\documentstyle[aps,epsfig,preprint]{revtex}

\tighten
\begin{document}
\title{Light charged particle evaporation from hot $^{31}$P nucleus at E$^{*}$ $\sim$ 60 MeV}

\author{D. Bandyopadhyay,  C. Bhattacharya,
K. Krishan, S. Bhattacharya, S. K. Basu}

\address{ Variable Energy Cyclotron Centre, 1/AF Bidhan Nagar,
Kolkata - 700 064, India}

\author{A. Chatterjee, S. Kailas, A. Shrivastava, K. Mahata}

\address{Nuclear Physics Division, Bhabha Atomic Research Centre,
Mumbai - 400 085, India}


\maketitle

\begin{abstract}

The inclusive energy spectra of light charged particles, such as, $\alpha$, p, d and t,
evaporated from hot $^{31}$P nucleus at excitation energy E$^{*}$ $\sim$ 60 MeV,
have  been  measured at various angles. The compound nucleus  $^{31}$P  has
been populated using two different entrance channel configurations; {\i.e.,}  
$^7$Li (47 MeV) + $^{24}$Mg and $^{19}$F (96 MeV) + $^{12}$C reactions,
leading to same excitation energy of the compound system. 
It  has  been observed that the spectra obtained
in the $^7$Li (47 MeV) + $^{24}$Mg reaction follow the standard statistical
model prediction with a
spherical configuration of the compound nucleus. But, the spectra obtained
in the $^{19}$F (96 MeV) + $^{12}$C reaction deviate 
from similar predictions of the statistical
model both on higher as well as on lower energy sides. Considerable
deformation was required to be incorporated in the calculation  
in order to reproduce the measured energy spectra in
this case. Dynamical trajectory model calculations were not found to play
any significant role in explaining the differences in behaviour 
between the two cases under study. 
The observed discrepancy has been attributed to the difference in the
angular momentum distributions of the compound nuclei 
formed in the two reactions.
\end{abstract}

\pacs{24.60.Dr, 25.70.Gh}

\vspace{0.2 cm}
\section{INTRODUCTION}
\label{sec1}

In recent years, there has been a lot of interest to study the statistical
properties of hot rotating nuclei   produced  
through heavy ion fusion reactions. High excitation energy
favours the  nucleus to deexcite through  statistical emission
of several light particles followed by  $\gamma$ decay.
The deexcitation process
is governed by the available phase spaces, {\it i.e.}, the nuclear level
densities  of both the parent and daughter nuclei, 
as a function of excitation energy, angular momentum and deformation.
The evaporated light charged particle spectra in
conjunction with the statistical model calculations can be used to extract
the statistical properties of the hot, rotating nuclei.

However, recent studies  reveal  that, at high
excitation energy and angular momentum, the  energy spectra of evaporated
light charged particles, are  no  longer  consistent with  the  standard statistical
model predictions. A large number of experiments have been performed in last
few years to study this anomaly over a  wide range of compound nuclear masses
(A$_{CN}$) in the range $\sim$60--170, for example \cite{5b,2,2a,3,3a,5,5a,1,5c,1a,1b,1c},  
and in  most  cases (with the exception of \cite{5b}), 
it has been found that the statistical model fails to
explain the  observed  light  charged  particle energy spectra. Choudhury
et al \cite{2} have measured the $\alpha$-particle energy spectra in the
reaction 214  MeV  $^{32}$S  on $^{27}$Al and observed that the
measured spectra deviate considerably from the  standard statistical
model predictions.  La Rana  et  al
\cite{2a}  have measured $\alpha$-particle spectra in the reaction
190 MeV $^{40}$Ar  on $^{27}$Al  and have  seen  similar  discrepancy. The
authors argued that in the statistical model calculation, the emission
barrier is  considered as
the fusion barrier of the two partners in the exit channel assuming their
spherical configuration. It may be a good approximation at low excitation
energy and angular momentum. But, when the nucleus is at high excitation
energy and rotating with high spin, it is likely to be deformed in shape.
Hence, the emission barrier will be lower if one considers that the
$\alpha$-particles are emitted preferentially along the larger axis. On the
contrary, it has been claimed  \cite{3,3a} that, only deformation, unless
it is  unrealistically  large,  cannot  affect  the  $\alpha$-particle
emission  barrier  much; however,
the anomaly may be well explained by incorporating  
spin  dependent  level  density  in  the   standard statistical  model
prescription  to  modify the yrast line keeping the emission barrier 
unchanged. Recently, Charity et al \cite{5a} have measured
light charged particle spectra emitted in  the  reactions  $^{64}$Ni  on
$^{100}$Mo  and  $^{16}$O  on $^{148}$Sm. They have shown  that the
measured  spectra  deviate from  the statistical model  predictions at incident
energies above $\sim$5 MeV/nucleon, and that 
different set of parameters are required to reproduce the data for different types  
of light charged particles, such as, $\alpha$, proton, deuteron and triton.
Moreover, magnitude of the deviation  has been found to depend on the
excitation energy as well as the entrance channel. 
Intuitively,  the entrance channel dependence may affect the
shapes  of the light charged particle spectra 
if the equilibrating system remains  deformed  over a
time scale comparable to the mean lifetime of  charged particle
emission. Recent measurements of  $\alpha$-particle energy
spectra for the reactions  83, 97 MeV $^{12}$C on $^{45}$Sc \cite{1b}
and 110MeV $^{16}$O on $^{54}$Fe  \cite{1a}
have also indicated that the role of entrance channel dynamics 
should be carefully looked into.

It is thus apparent  that there exists some discrepancy between the
experimentally measured spectra and the corresponding predictions of
standard statistical model, which is prominent for higher spin and for
heavier systems (A$\ge$60). However, due to lack of experimental data
situation is not clear for the lighter systems at relatively lower spin.
This prompted us to  explore  the scenario for  light system, such as,
$^{31}$P, in the incident energy range  below 5  MeV/A. 
In the present work, we have populated
the compound nucleus $^{31}$P at  excitation energy  $\sim$
60 MeV  through  different entrance channels, i.e., 96 MeV $^{19}$F on
$^{12}$C and 47 MeV $^7$Li on $^{24}$Mg and have measured the energy 
and angular  distributions of all  light charged particles, (i.e., $\alpha$, p, d and t). 
Since the two reaction channels
populate the same compound system with different angular momentum distributions
(l$_{cr}$ $\sim$ 21 $\hbar$ and 16 $\hbar$, respectively), it offers
an opportunity to study the angular momentum dependence in the evaporative
decay cascade. Apart from that,  entrance channel configurations may also
affect the evaporative  decay  if the nucleus lives in
the dinuclear form for a long time and evaporates during this period. But
our previous studies on the intermediate mass fragment emission  did not
reflect the existence of any such long-lived dinuclear
configuration\cite{7,7b} of the $^{31}$P compound nucleus.  
Earlier, we have measured  also $\alpha$-particle spectra  in coincidence  with  the
evaporation  residues  in  the  96 MeV $^{19}$F on $^{12}$C reaction
\cite{5c} and have observed that the measured spectra deviate from the
statistical model predictions both at low as well as at high energy sides. Assuming  a
deformed configuration of the compound nucleus, the full energy spectra
could be explained satisfactorily. Interestingly, it has been observed in earlier works that, 
different amount of deformations are required  to
explain  different sets of light charged particle data. It
directly points to the inadequacy of the standard statistical
model to explain the behaviour of light charged particle emission spectra with increasing 
spin and excitation energy.  In the present work, we would like to  
explore systematically the scenario of light charged
particle emission further for a lighter system at moderate spin and
excitation energy. 

The  paper  has  been  arranged  as follows. The experimental details
alongwith the experimental results are described in Sec.~\ref{sec2}.  The
predictions  of the statistical model for $^{31}$P nucleus have
been discussed in Sec.~\ref{sec3}. Finally the  summary  and conclusions are
presented in Sec.~\ref{sec4}.

\section{EXPERIMENTS AND RESULTS}
\label{sec2}

 The  experiments  were  performed  at the Bhabha Atomic Research Centre -
Tata Institute of  Fundamental  Research  14  UD  pelletron  accelerator
laboratory,  Mumbai,  with  96  MeV  $^{19}$F  beam on 125$\mu$g/cm$^2$
self-supporting $^{12}$C target and 47 MeV $^7$Li beam on 275 $\mu$g/cm$^2$
self-supporting $^{nat}$Mg target. The beam  current  was  typically 20-80
nA.  The
light  charged  particles  have  been  detected  in  three  solid  state
telescopes. The
telescopes  (T$_1$,  T$_2$,  T$_3$) were of two elements, consisting of 100$\mu$m,
45$\mu$m, 40$\mu$m $\Delta$E Si(SB) and 5  mm,  5  mm,  2  mm  Si(Li)  E
detectors  respectively.  Typical solid angles were 1.5 msr, same for all
the three telescopes. Analog signals from the detectors  were  processed
using  the  standard  electronics  before  being fed to the computer for
on-line data acquisition.

 The   telescopes   were  calibrated  with  the  radioactive  $^{228}$Th
$\alpha$-source. Typical energy resolutions (FWHM) obtained were 1.12$\%$
for T$_1$ and 1.13$\%$ for T$_2$ and T$_3$  respectively.  The low energy
cutoffs  thus  obtained  were  typically  3  MeV  for  proton, 5 MeV for
deuteron, 5 MeV for triton and 11 MeV for $\alpha$ in T$_1$. In T$_2$  and 
T$_3$  cutoffs  were  1 MeV for proton, 3 MeV for deuteron, 3 MeV for triton
and 7 MeV for $\alpha$-particles.

\begin{figure}
\centering
\epsfig{file=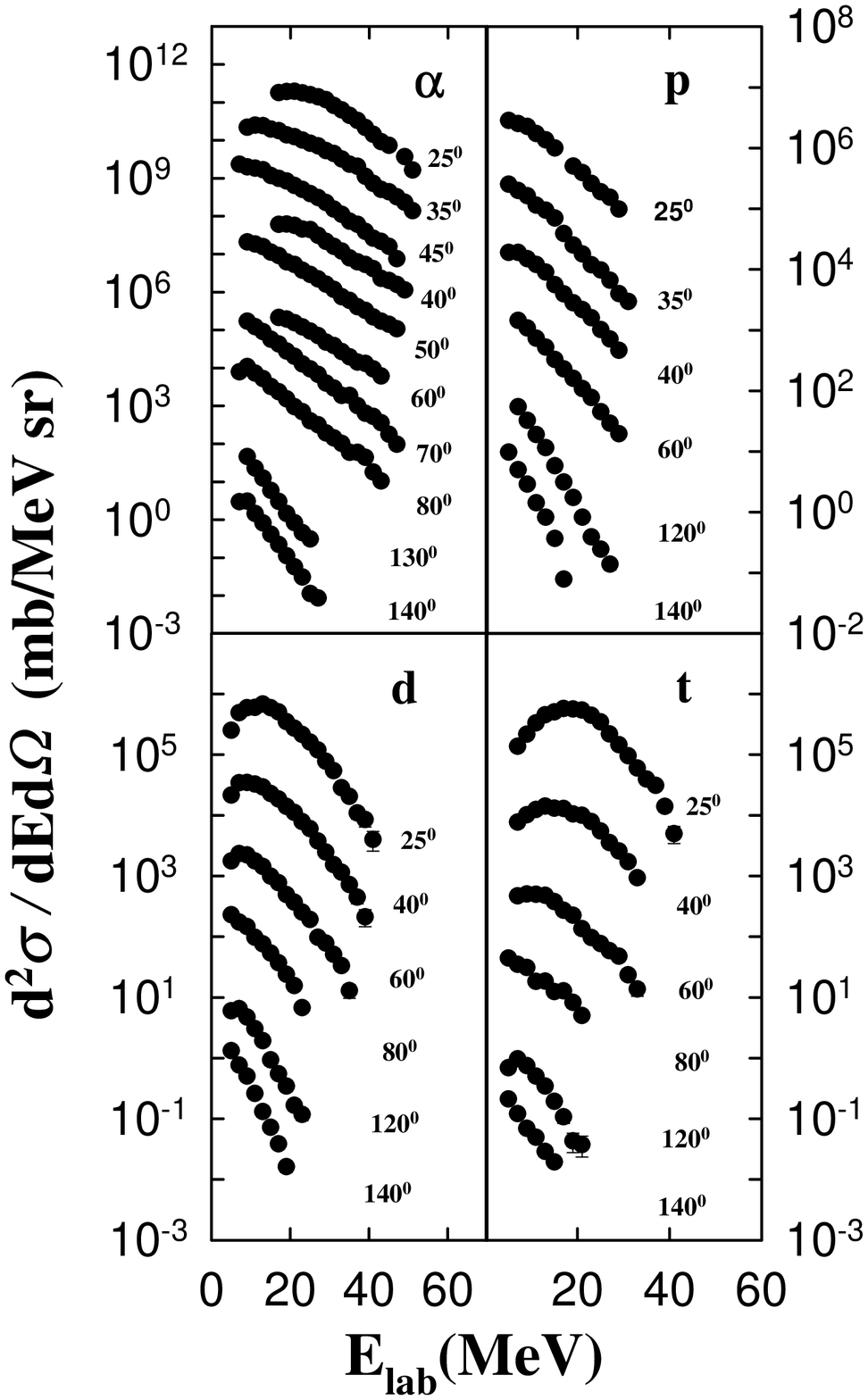,width=12cm}

\caption{inclusive  (filled circles) light charged particle energy spectra
obtained in the $^7$Li (47 MeV) + $^{24}$Mg reaction at different laboratory
angles. For each curve cross-sections at highest observed angle is multiplied
by 10$^{\circ}$ and the multiplication factor subsequently increases by a
factor of 10 for each of the successive curve as the angle increases.}
\label{fig1}
\end{figure}

\begin{figure}
\centering
\epsfig{file=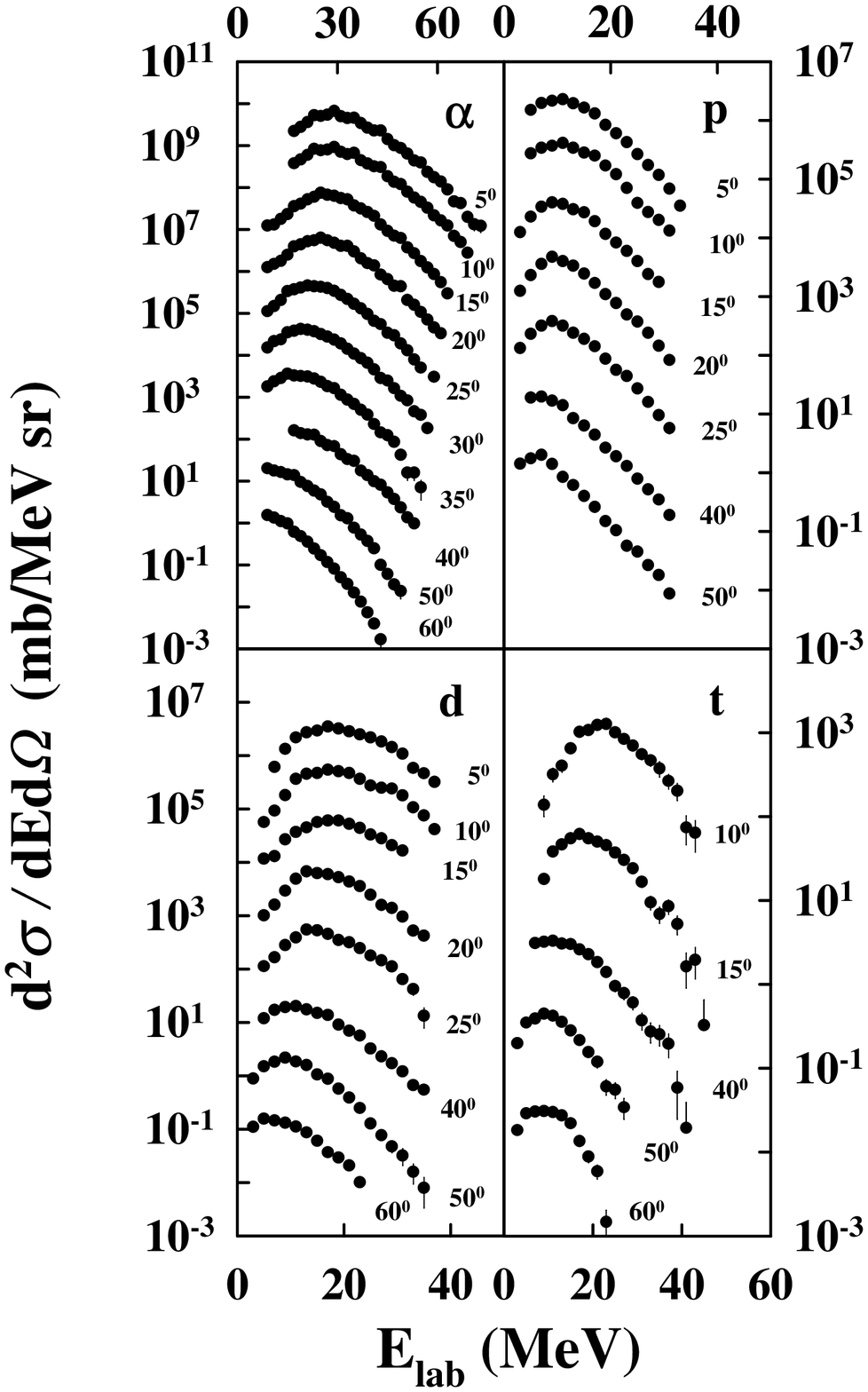,width=12cm}
\caption{Same as fig. 1 for the $^{19}$F (96 MeV) + $^{12}$C reaction.}
\label{fig2}
\end{figure}

Inclusive energy distributions for $\alpha$-particles, proton, deuteron
and triton have  been  measured  in  an  angular  range of 25$^{\circ}$ -
140$^{\circ}$ in case of $^7$Li (47 MeV) + $^{nat}$Mg reaction and in a
range of 5$^{\circ}$-60$^{\circ}$ in case of $^{19}$F (96 MeV) + $^{12}$C
reaction. In  the centre of mass (c. m.) frame it covers up to $\sim$
140$^{\circ}$ in both the cases. Figs.~\ref{fig1}-\ref{fig2} show the
measured energy spectra for the light  charged particles  at  different
laboratory angles respectively for the two reactions. The shape of the
energy spectra is similar for all the measured angles indicating  the  fact
that  the thermal equilibrium has been achieved in both the cases before
the particle emission.

\begin{figure}
\centering
\epsfig{file=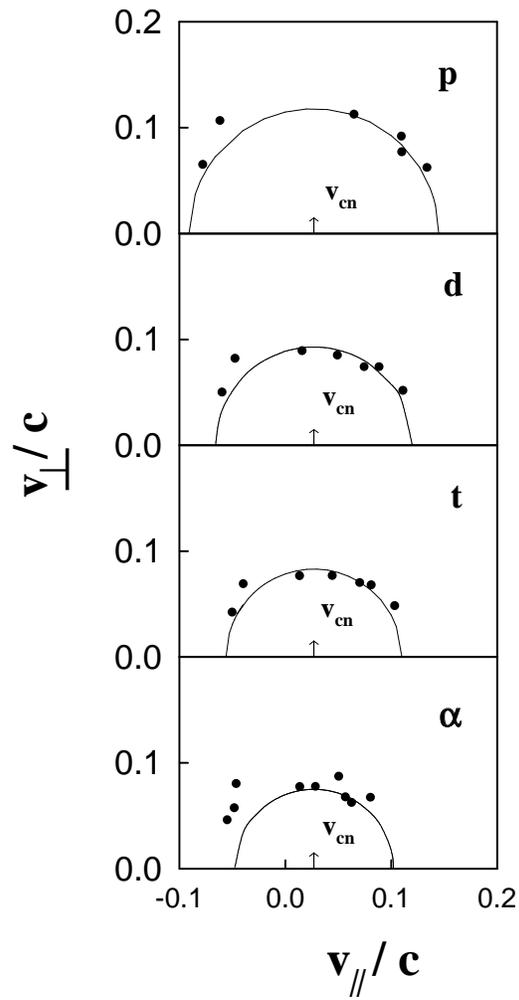,width=12cm}
\caption{Average velocities of the light charged particles, obtained in
the $^7$Li (47 MeV) + $^{24}$Mg reaction, as a function
of velocities parallel ($v_{\parallel}$) and perpendicular ($v_{\perp}$)
to  the beam direction. The arrow indicates the velocity of the compound
nucleus.}
\label{fig5}
\end{figure}

\begin{figure}
\centering
\epsfig{file=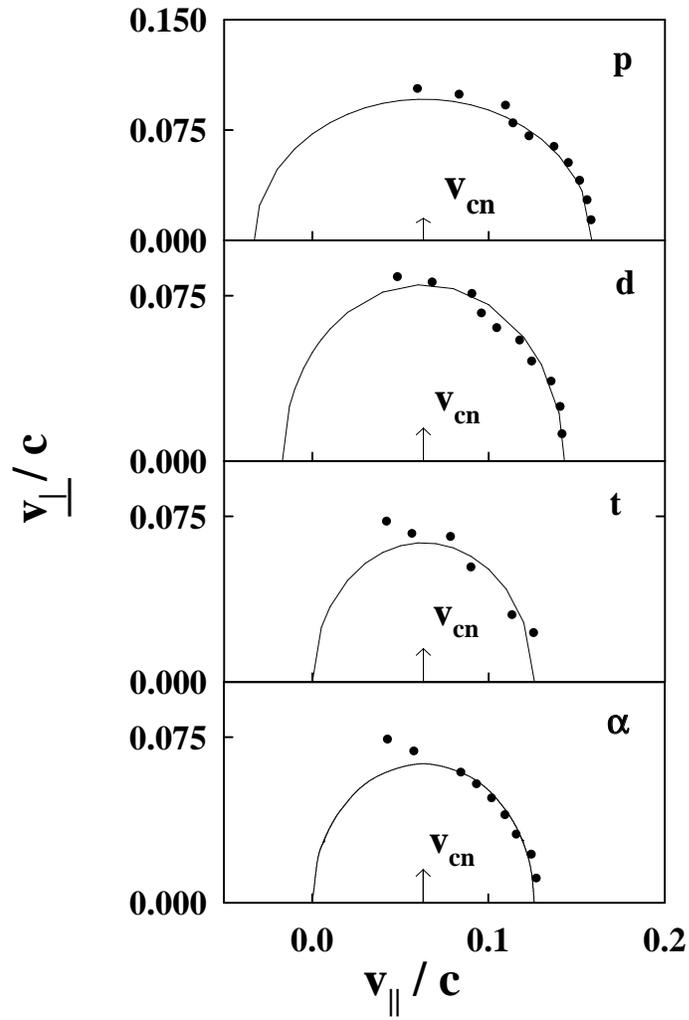,width=12cm}
\caption{Same as fig. 3 for the $^{19}$F (96 MeV) + $^{12}$C reaction.}
\label{fig6}
\end{figure}

The  average  velocities  of  the observed light charged particles have
been plotted  in the parallel ($v_{\parallel}$) and perpendicular
($v_{\perp}$) velocity plane ( Figs.~\ref{fig5}-\ref{fig6}) for the
$^7$Li (47 MeV) + $^{24}$Mg
and $^{19}$F (96 MeV) + $^{12}$C  reactions. It has been observed, that
for all particles, the average velocities fall
on  a  circle  around  the  compound  nucleus  velocities ($v_{cn}$).
It implies  that  the average velocities (as well as the
average kinetic energies) of the light charged particles are independent
of the centre of mass emission angles in both the reactions. It means
that the energy relaxation is complete and the light charged  particles
are  emitted  from  a fully equilibrated source moving with the velocity
$v_{cn}$.  The deviations in average velocity of the light charged 
particles at backward angles ($\theta_{lab}\ge40^{\circ}$) are  due to
large  energy  cutoffs  in  the  energy spectra.

\section{DISCUSSION}
\label{sec3}

\begin{figure}
\centering
\epsfig{file=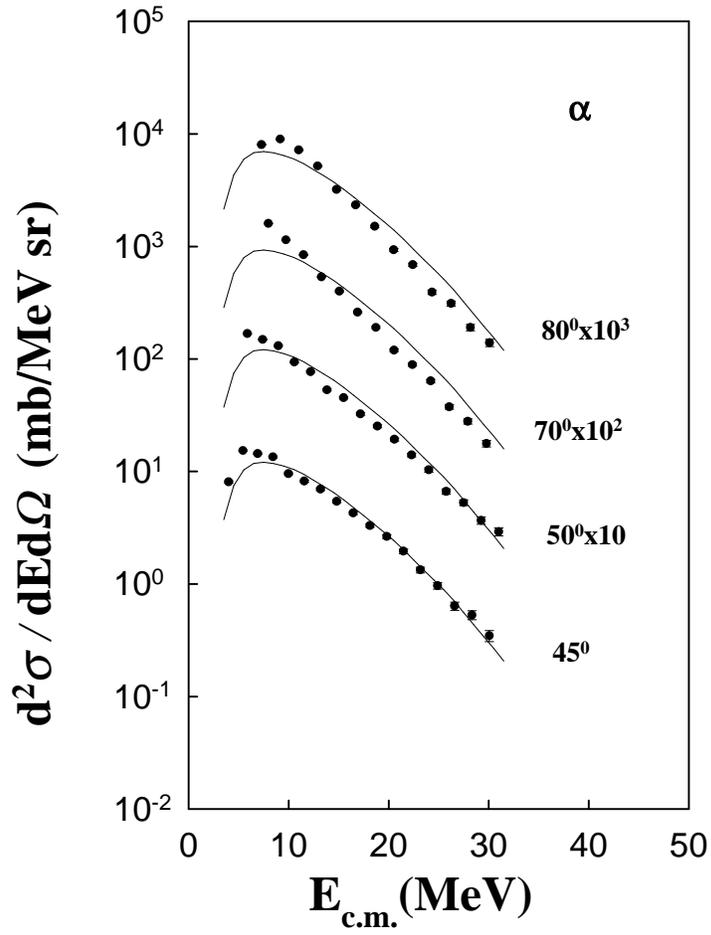,width=12cm}
\caption{inclusive  (filled circles) $\alpha$-particle energy spectra,
obtained in the $^7$Li (47 MeV) + $^{24}$Mg reaction, in
centre-of-mass frame measured at different laboratory angles. Solid
lines (normalized at the peak of the measured data) are the prediction
of CASCADE calculations with the standard
value of the radius parameter, $r_0$.
({\it see text}).}
\label{fig7}
\end{figure}

\begin{figure}
\centering
\epsfig{file=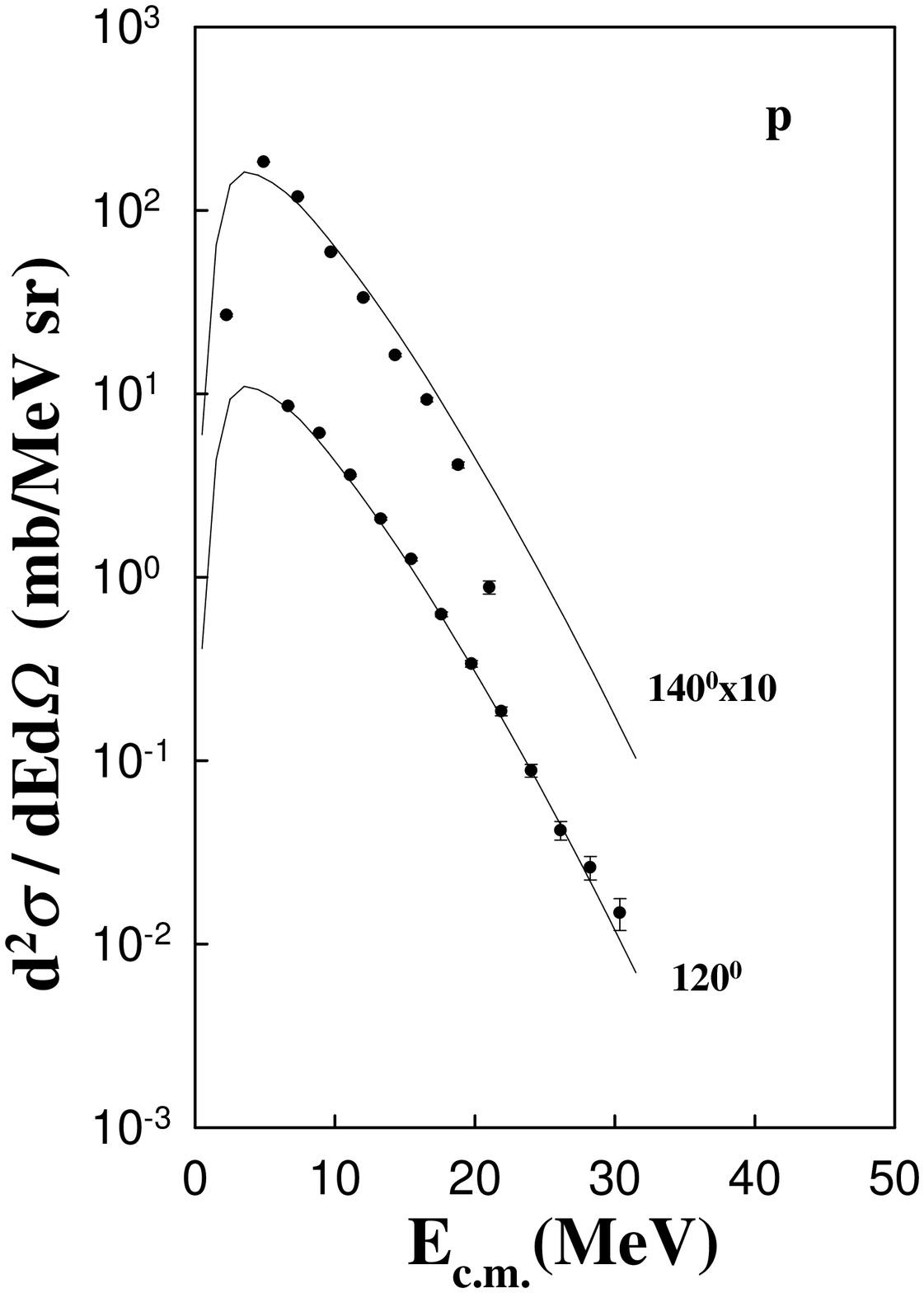,width=12cm}
\caption{Same as fig. 5 for protons.}
\label{fig8}
\end{figure}

\begin{figure}
\centering
\epsfig{file=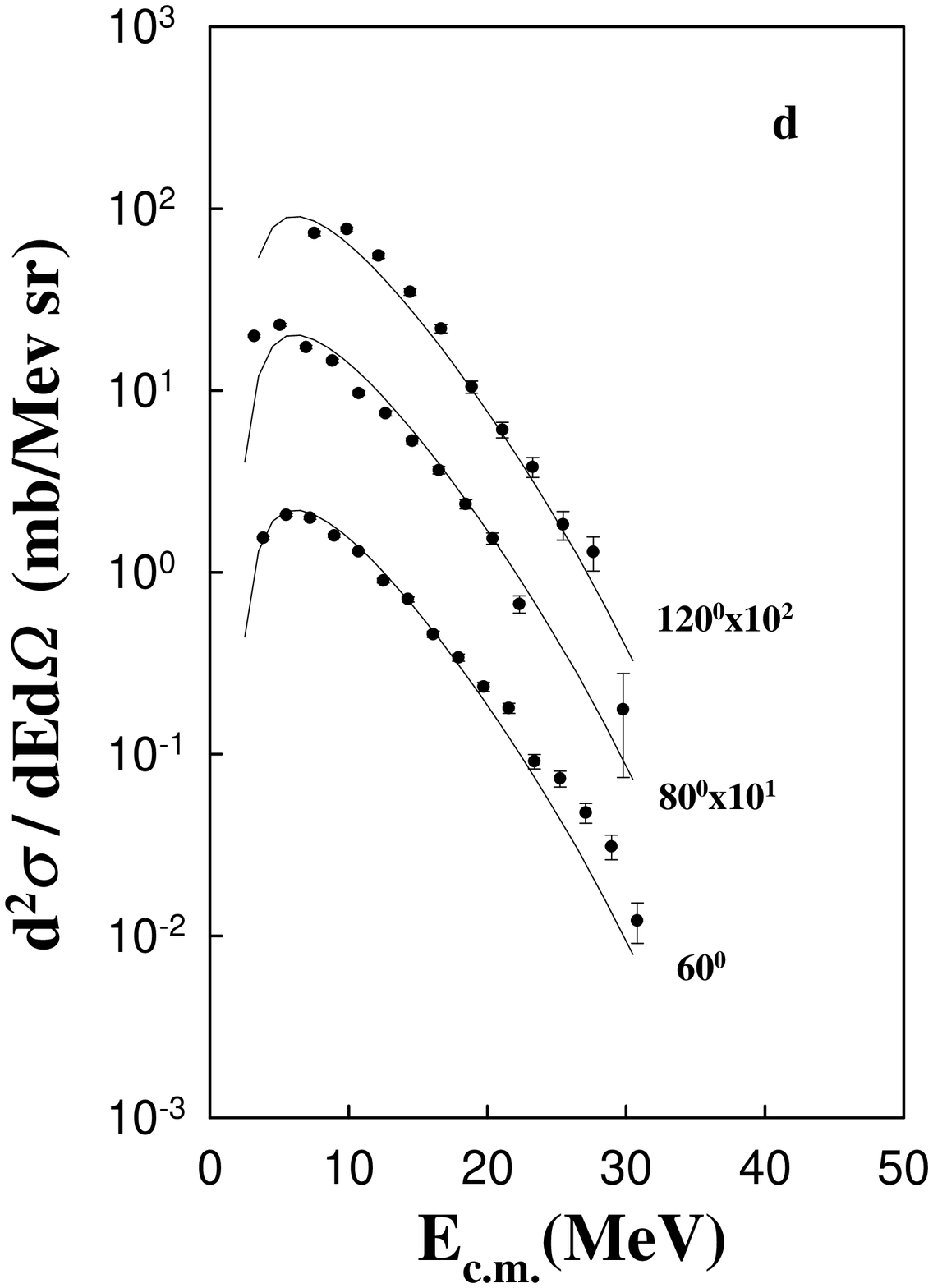,width=12cm}
\caption{Same as fig. 5 for deuterons.}
\label{fig9}
\end{figure}

\begin{figure}
\centering
\epsfig{file=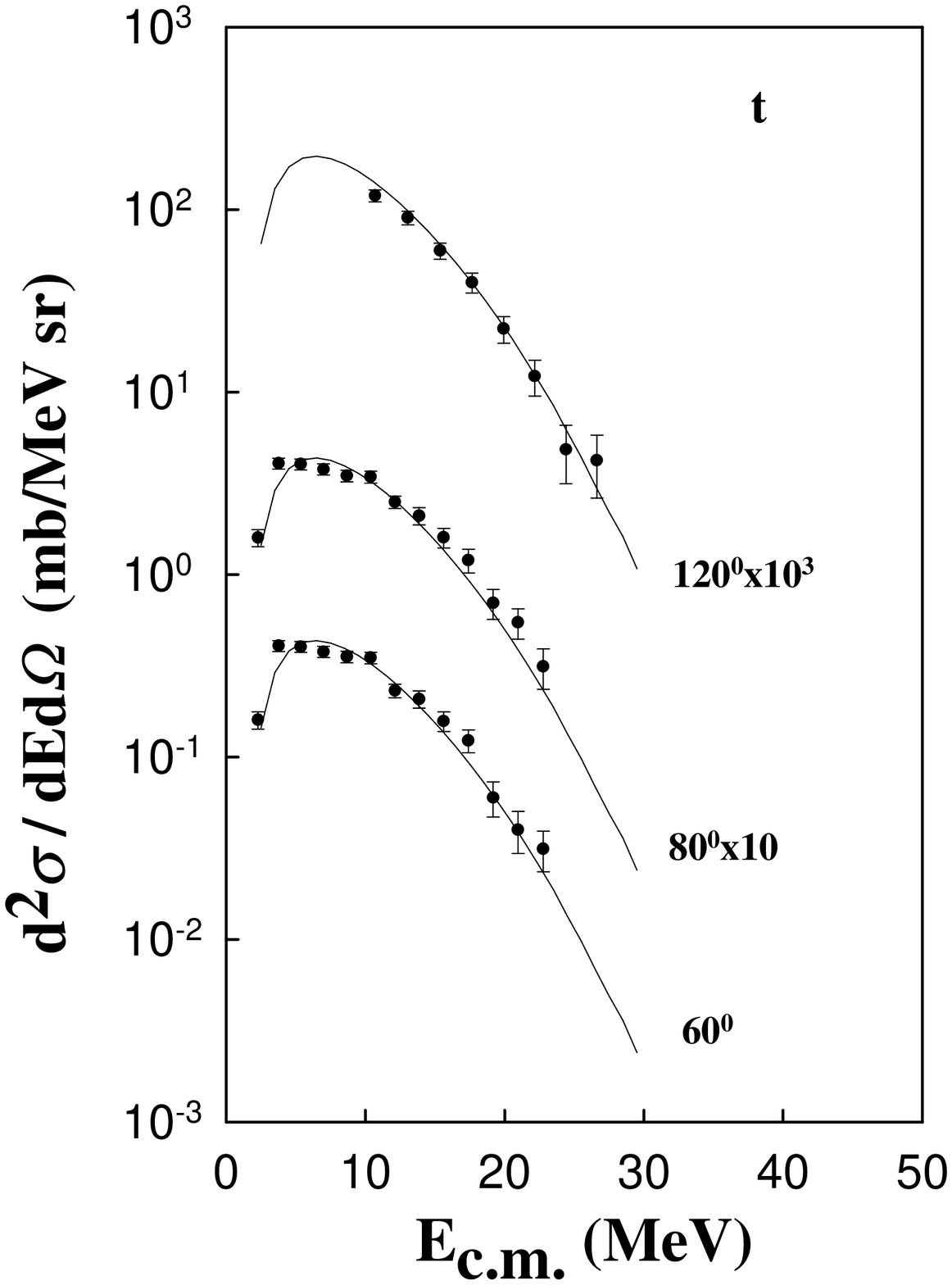,width=12cm}
\caption{Same as fig. 5 for tritons.}
\label{fig10}
\end{figure}

The experimental data have been compared with  the  predictions of standard
statistical  model  calculation using the code CASCADE \cite{6}.
Figs. ~\ref{fig7} - \ref{fig10} show the centre of mass energy spectra
for the various light charged particles, such as $\alpha$, p, d and t,  
observed in the $^7$Li (47 MeV) + $^{24}$Mg reaction.
The solid lines represent the predictions of the statistical model code
CASCADE. The optical  model parameters used  for
calculating the  transmission  coefficients  
were taken from  \cite{6a} for {\it p, d, t}, and from  \cite{6c} for 
$\alpha$-particles.  The  critical
angular   momentum  for  fusion,  $l_{cr}$,  is  calculated using Bass model
as 16$\hbar$ \cite{6b}. Sharp cut-off angular momentum distribution has
been assumed. The
radius parameter was taken to be 1.29 fm following  P\"{u}hlhofer  et  al
\cite{6}. The angular
momentum dependent deformation parameters ($\delta_1$ and $\delta_2$ as
described later) were set equal to zero. From the figures, it is observed
that the theoretical  predictions  match  well  with the experimental
results, indicating that all the light charged particles are
evaporated from a nearly spherical $^{31}$P compound nucleus . 
                  
\begin{figure}
\centering
\epsfig{file=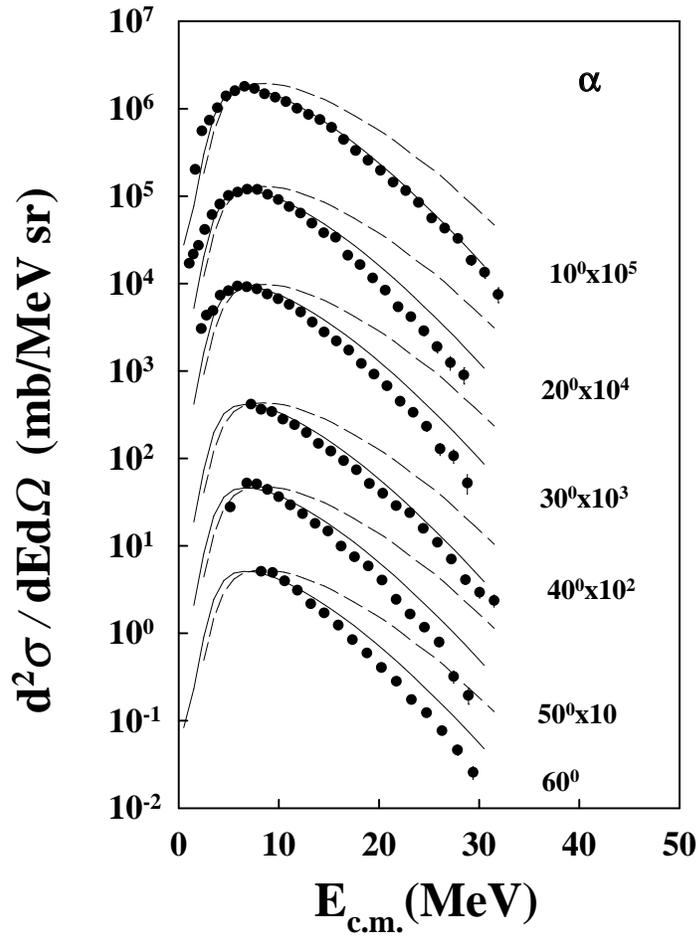,width=12cm}
\caption{inclusive  (filled circles) $\alpha$-particle energy spectra,
 obtained in the $^{19}$F (96 MeV) + $^{12}$C reaction, in
centre-of-mass frame measured at different laboratory angles. Curves are
the predictions of CASCADE calculations with the standard (dashed  curves)
and  the  modified  (solid curves) values of the radius parameter, $r_0$
({\it see text}). Curves are normalized at the peak of the measured data.}
\label{fig11}
\end{figure}

\begin{figure}
\centering
\epsfig{file=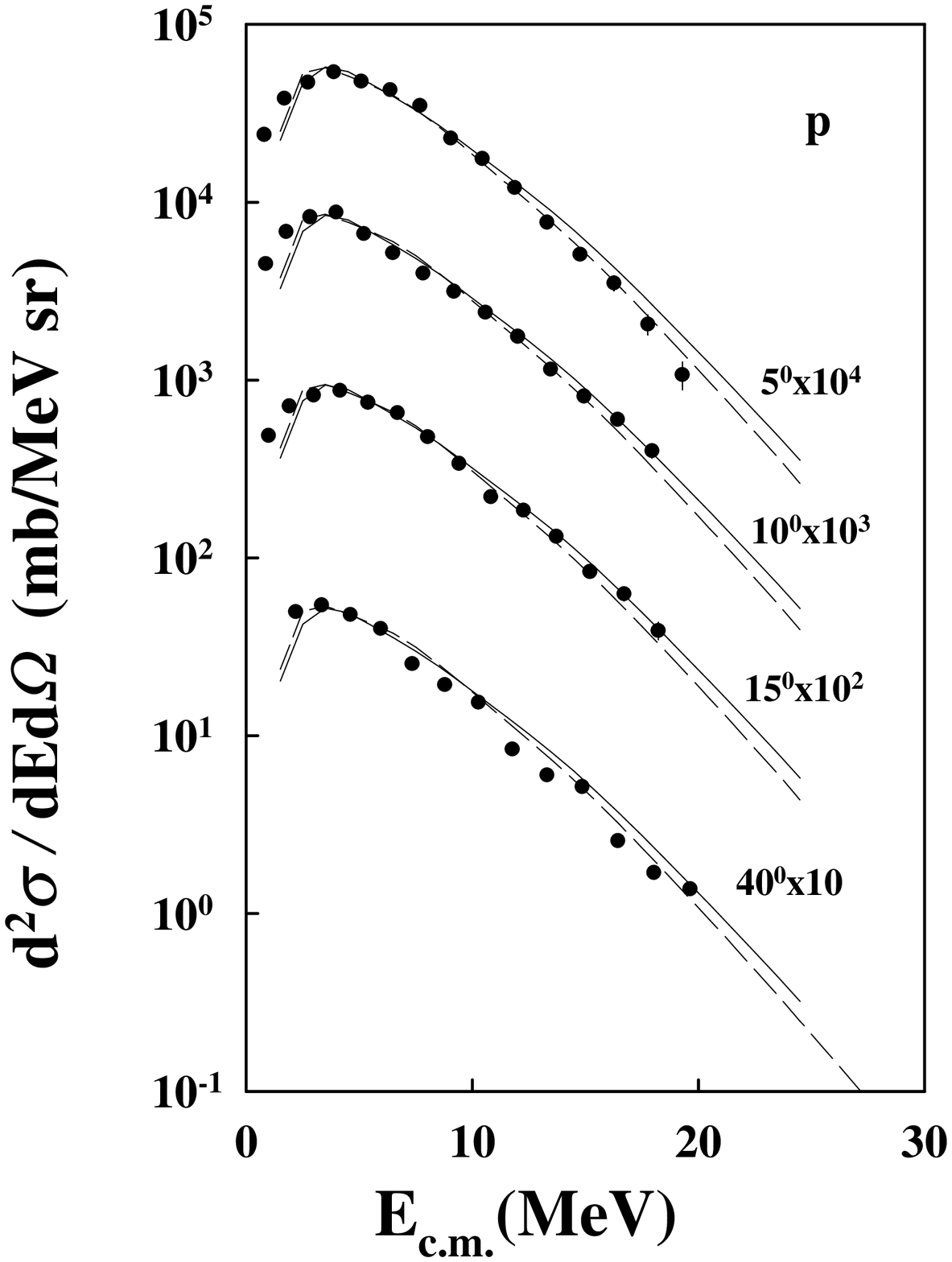,width=12cm}
\caption{Same as fig. 9 for protons.}
\label{fig12}
\end{figure}

\begin{figure}
\centering
\epsfig{file=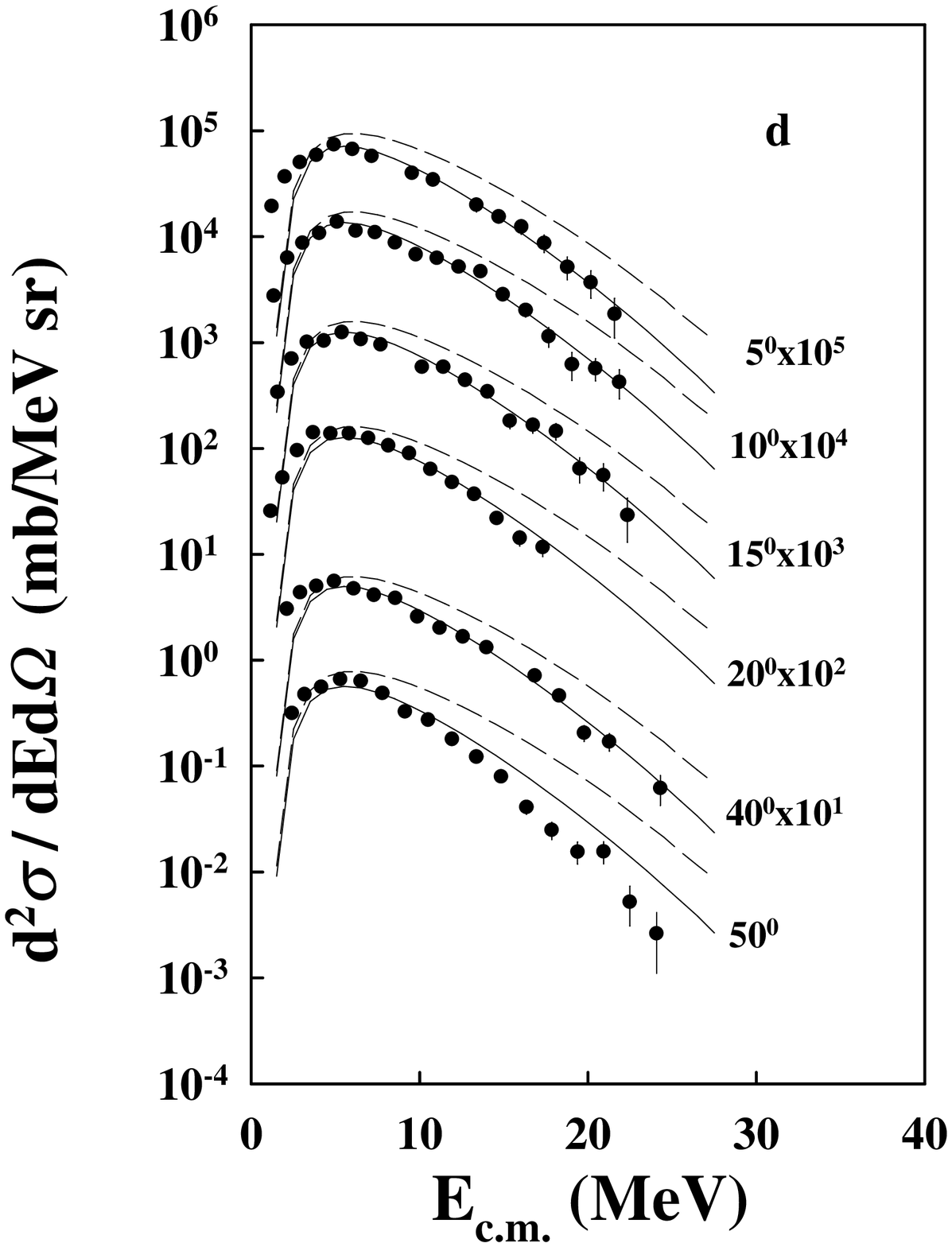,width=12cm}
\caption{Same as fig. 9 for deuterons.}
\label{fig13}
\end{figure}

\begin{figure}
\centering
\epsfig{file=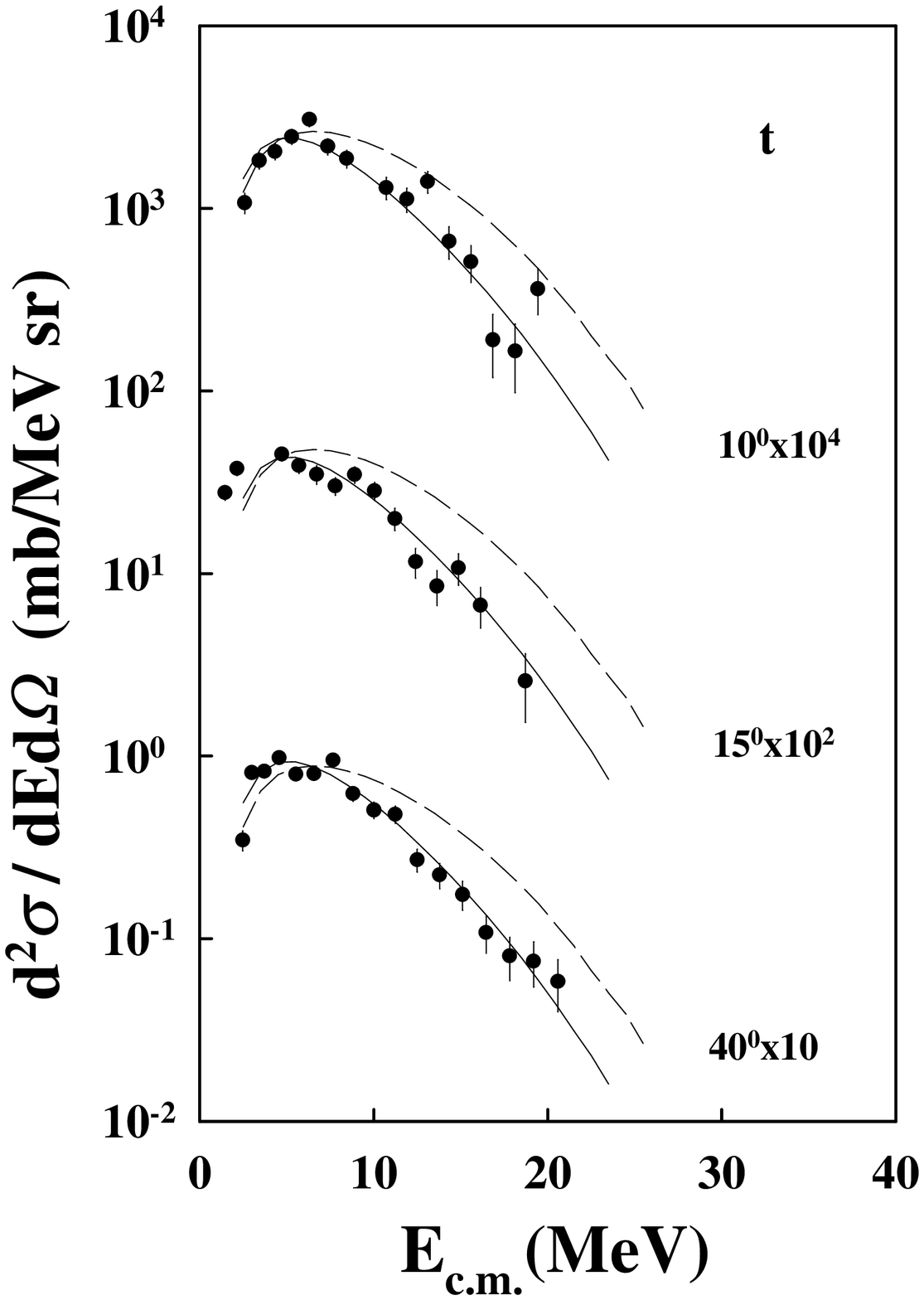,width=12cm}
\caption{Same as fig. 9 for tritons.}
\label{fig14}
\end{figure}

However, in case of $^{19}$F (96 MeV) + $^{12}$C  reaction, the statistical
model calculation using
the code CASCADE, with the standard parameter set, fails (the dashed lines) 
to predict the shape of the experimental spectra. Figs. 
~\ref{fig11} - \ref{fig14} show the observed (filled circles) and calculated
(solid and dashed lines) centre of mass energy spectra
for the various light charged particles emitted in the $^{19}$F (96 MeV)
+ $^{12}$C reaction. The figures indicate deviation of the experimental
data from the standard statistical model predictions (dashed lines) at
low as well as at high energy sides of the spectra. 

According to  the  CASCADE  calculations, the lower energy part of
the spectra is controlled by the transmission coefficients.  
For the  calculation of transmission coefficients, 
we have used the same sets of optical model parameters 
as in the case of  $^7$Li (47 MeV) + $^{24}$Mg reaction,
({\it i.e.}, the optical model parameters  from \cite{6c} for alpha
particles, and  from \cite{6a} for proton, deuteron and triton). A deformed
configuration of the compound nucleus has been assumed in the calculation.
This is realized by varying only the radius parameter r$_{\circ}$. The
optimum value of r$_{\circ}$ is found to be $\sim$ 1.56 fm which is
consistent with our earlier work \cite{5c}. The increased radius parameter
lowers  the emission barrier and hence
enhances the low energy yield by increasing the transmission coefficients.

The higher energy part of the spectra is controlled by the level density
$\rho(E,j)$, which for a given angular momentum $j$ and excitation energy
$E$ is defined as \cite{6},

\begin{equation}
\rho(E,j) = \frac{2j+1}{12}a^{1/2}(\frac{\hbar^2}{2{\cal J}_{eff}})^{3/2}\\
  \frac{1}{(E+T-\Delta-E_j)^2} \exp[2[a(E-\Delta-E_j)]^{1/2}]
\label{lev}
\end{equation}

where $a$ is the level density parameter, which is taken to be $A/8$ for
the  present calculation, $T$ is the thermodynamic temperature, $\Delta$
is the pairing correction, and E$_j$ is the rotational energy which  can
be  written in terms of the effective moment of inertia ${\cal J}_{eff}$
as

\begin{mathletters}
\begin{equation}
E_j =\frac{ \hbar^2}{2 {\cal J}_{eff}}j(j+1).
\label{erot}
\end{equation}

In  hot  rotating nuclei, formed in heavy ion reactions, the effective
moment of inertia may be spin dependent and may be written as,

\begin{equation}
{\cal J}_{eff}= {\cal J}_0 \times (1+\delta_1j^2+\delta_2j^4),
\label{jeff}
\end{equation}

where, the rigid body moment of inertia,  ${\cal J}_0$, is given by,

\begin{equation}
{\cal J}_0 = \frac{2}{5} A^{5/3} r_{\circ}^2.
\end{equation}
\end{mathletters}

Here,  A  is  the mass number and r$_\circ$ is the radius parameter. Non
zero values of the  parameters  $\delta_1$,  $\delta_2$  introduce  spin
dependence  in  the  effective  moment  of  inertia (Eq.  \ref{jeff}).    
The relation (Eq. \ref{erot}) defines  a
region  in  the  energy, angular momentum plane (E-j plane) of
allowed levels and a region where no levels occur. The region is bounded
by the yrast line defined by the lowest energy state that occurs  for  a
given angular momentum. It has been observed in the previous experiments
\cite{3a,5c} that the yrast line has to be modified to explain the experimental
data. Different attempts have been pursued using different tools to modify the
yrast  line,  such  as, radius parameter $r_\circ$, deformation parameters
$\delta_1$, $\delta_2$ and the angular momentum j.

However,  we  have  observed that the variation of a single input parameter,
$r_\circ$, in the code CASCADE, is sufficient to take care of the low energy
as  well  as  the high energy side of the spectra\cite{5c}. So, in the
present calculation, we varied the parameter $r_\circ$  only,  and  kept
the    spin    dependent    parameters    $\delta_1$,    $\delta_2$   in
Eq.~(\ref{jeff}) equal to zero, to reproduce the experimental  spectra.
It  is  clear from Figs.~\ref{fig11} - ~\ref{fig14} that the calculated
spectra (solid lines) agree quite well with the experimental spectra both in  the  low
energy  as  well as in the high energy regions.

\begin{figure}
\centering
\epsfig{file=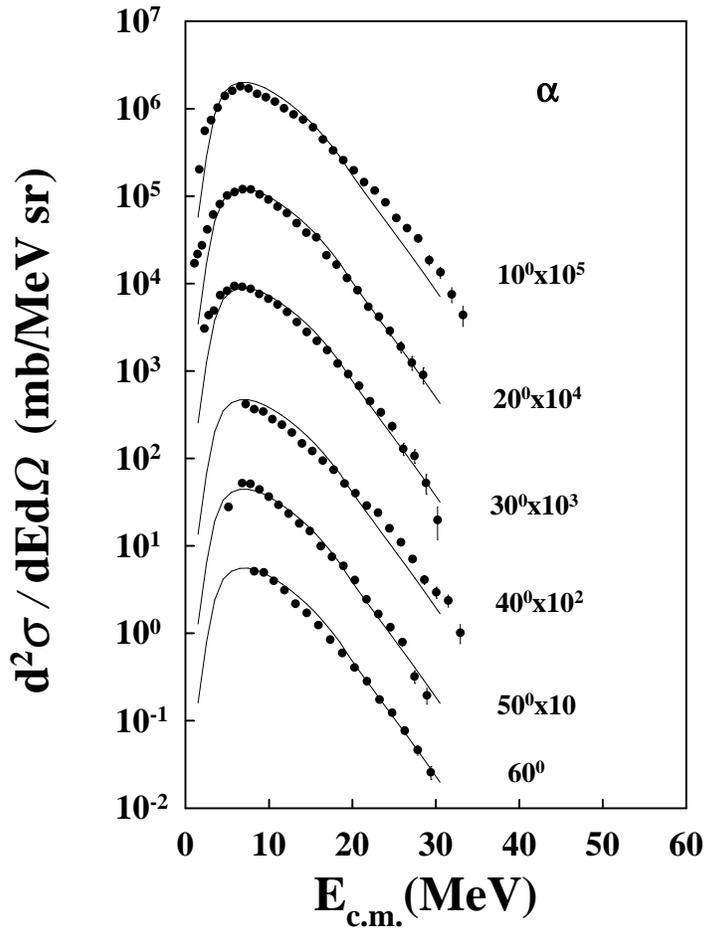,width=12cm}
\caption{Inclusive c.m. $\alpha$-particle energy spectra at different laboratory
angles, obtained in the $^{19}$F (96 MeV) + $^{12}$C reaction. Solid lines
(normalized at the peak of the measured data) are the predictions of CASCADE
with $\delta_1$ = 2.8 $\times$ 10$^{-3}$ and $\delta_2$ = 2.5 $\times$ 10$^{-7}$
({\it see text}).}
\label{fig16}
\end{figure}

For the sake of completeness, we  also attempted to interpret the data using
the other alternative; {\it i.e.}, by introducing  angular momentum dependence in the
effective moment of inertia through the angular momentum dependent 
deformation parameters, i.e., $\delta_1$, $\delta_2$  of
Eq.~\ref{jeff}.  The best fit values $\delta_1$ and $\delta_2$  obtained from
fitting the data were
$\delta_1 = 2.8 \times 10^{-3}$ and  $\delta_2  =  2.5  \times  10^{-7}$.
A representative theoretical  prediction with the above best-fit parameters have 
been shown in Fig.~\ref{fig16} along  with  the  respective experimental  data  
for  the $\alpha$-particles
emitted in   $^{19}$F (96 MeV) + $^{12}$C reaction.  It  is  clearly seen that
with the above values  of   $\delta_1$ and $\delta_2$,
the  predicted spectra  match well with the experimental ones. It is further
evident (comapring Fig.~\ref{fig16} with Fig.~\ref{fig11}) that both the 
prescriptions (either changing  $r_\circ$  only or varying  $\delta_1$ and 
$\delta_2$) yield nearly equivalent result, at least for the system considered here.

In order to probe more deeply into the matter,  we   have
estimated  an  average  radius  parameter (for non zero  values of $\delta_1$ and 
$\delta_2$) by  averaging over the angular
momentum distribution (sharp cut-off  triangular  distribution  has  been
assumed, upto  the  critical  angular
momentum 21$\hbar$) in the following way. 

\begin{equation}
r_{eff}^2=r_{\circ}^2\frac{\sum^{j_{cr}}_{\circ}(1+\delta_1j^2+\delta_2j^4)
(2j+1)}{\sum^{j_{cr}}_{\circ}(2j+1)}
\end{equation}

Here, r$_{\circ}$ is taken as 1.29 fm. j$_{cr}$ is the critical angular
momentum of the system. $\delta_1$ and $\delta_2$ values are as stated 
above. The effective radius parameter, r$_{eff}$, thus obtained is 1.65 fm
which  is  within 10$\%$  of  the  radius  parameter  we  have used in the
statistical model calculation to reproduce the experimental data.
Therefore, it can be said that the variation
of single parameter r$_{\circ}$ to fit the data is nearly equivalent
to the spin dependent level density prescription as described above,
at least for light compound systems involving lower spins.

For such light compound systems like $^{31}$P at low spins, the
present method has a distinct advantage. Here, both lower and
higher energy side of the spectra could be fitted by changing a single
parameter r$_{\circ}$, once for all. The increase of r$_{\circ}$
causes lowering of the barrier which enhances the transmission 
coefficient, thereby correcting the lower energy side of the spectra;
at the same time, it also causes a reduction of rotational energy
thereby changing the level density in right direction to explain the
high energy part of the spectra. On the contrary, in the standard
spin dependent level density prescription, one has to take care
of the low energy side of the spectra separately by introducing a 
reduced barrier. However it may be noted that, as the spin dependent 
deformation becomes more significant at higher spins,
the present method may not lead to  realistic values of r$_{\circ}$ 
for higher spins;  in such cases one may have to consider
the spin dependent deformation only. On the other hand, for lower
spin values, (for example, the compound nucleus
$^{31}$P formed in the reaction $^7$Li (47 MeV) + $^{24}$Mg with 
l$_{cr}$  $\sim$ 16 $\hbar$) the change in r$_{\circ}$ may not be
quite appreciable. This intuitively explains
why the standard statistical model predictions have been in
good agreement with the data in this case.

\begin{figure}
\centering
\epsfig{file=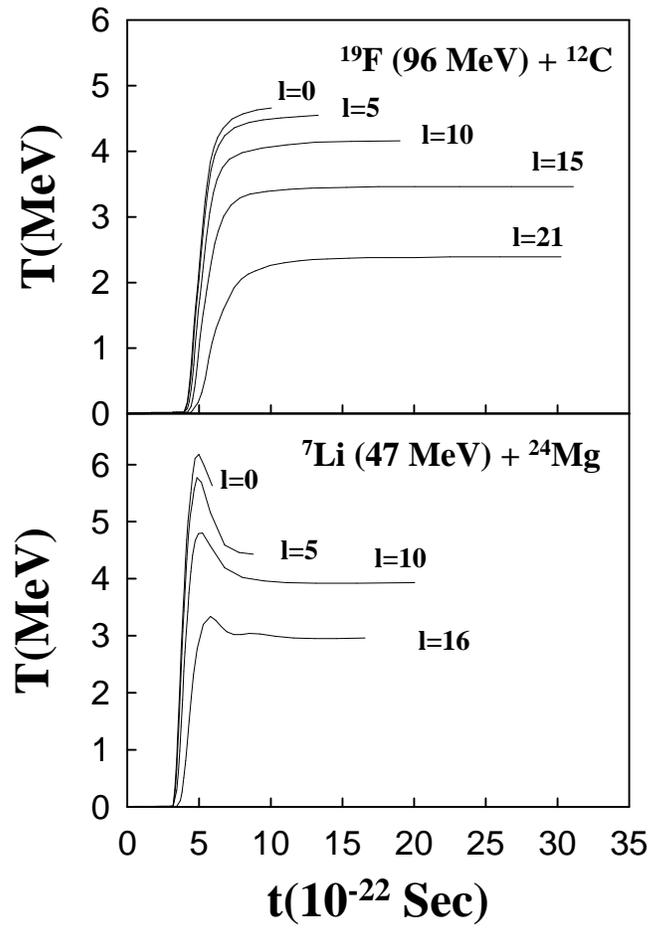,width=12cm}
\caption{Time evolution of the temperature (T) parameter as obtained from
HICOL calculations for the reactions studied.}
\label{fig15}
\end{figure}

We have also tried to estimate the effect of entrance channel dynamics
on light charged particle emission using the code HICOL \cite{5d}.
Fig.~\ref{fig15} shows the calculated temparature of the fused nuclei as a
function of time for various angular momentum  values  covering  the  range
of interest for the two reactions under study. From  Fig.~\ref{fig15}, it is 
observed that for both the cases, thermal equlibration of the system is achieved, on the
average,  within  8$\times$10$^{-22}$  second after the zero time. The zero time is
defined as the time when the participating  nuclei  begin  to  feel  the
nuclear  forces and deviate from their earlier coulomb trajectories. It
is also observed that the  temperature  decreases  with  increasing  
angular momentum. Decay time has been estimated to be $\sim$12$\times$10$^{-22}$
second, using the  code  PACE2,  which is comparable  to the formation time of
the nucleus as obtained from HICOL. This implies that the effect of entrance channel on
particle emission should be similar in both the reactions.
However, particle spectra observed in the two reactions are different, implying that  
the above difference   may not be due to the entrance channel dynamics.

\section{SUMMARY AND CONCLUSION}
\label{sec4}

We  have measured the inclusive light charged particle energy spectra  
in an angular range of 5$^{\circ}$-60$^{\circ}$ and
25$^{\circ}$-140$^{\circ}$
respectively in the $^{19}$F (96 MeV) + $^{12}$C and $^7$Li (47 MeV)
+ $^{nat}$Mg reactions.  It has been observed from the average
velocity plots in the $v_{\parallel}$ vs. $v_{\perp}$ plane that 
the average velocities of all  light charged
particles in both the reactions stated above fall on the circles centered
around their respective compound nuclear velocities, implying that these 
particles are emitted from  equilibrated compound nuclear sources. 
The data in the two reactions have been compared with the statistical model
calculations using the code CASCADE. It has been observed that the light
charged particle spectra measured in the $^7$Li (47 MeV) + $^{24}$Mg
reaction can be well explained in the framework of standard statistical
model calculation, whereas the light charged
particle spectra obtained in the $^{19}$F (96 MeV) + $^{12}$C  reaction 
deviate significantly from the similar calculations. 
The formation times as estimated by the semiclassical calculation
using the code HICOL and the decay times as estimated by PACE2 are comparable
in both the cases. 
Therefore, the discrepancy in the light charged particle
spectra obtained in the two reactions may not be  due to the
entrance channel effect as the estimated formation and decay times of the compound
nucleus in both the cases are of the same order. Rather, it may be attributed to the
difference in the angular momentum distributions of the compound nuclei formed in the two 
reactions considered here. 
A satisfactory description of the measured energy
spectra in the case of  $^{19}$F (96 MeV) + $^{12}$C
has been achieved by invoking a deformed configuration of  the
compound nucleus  through  the  modification  of the radius parameter
$r_\circ$ from 1.29 fm to 1.56 fm in the statistical model calculation.
The data could also be reproduced equally well by 
incorporating spin dependent level density formalism and the average
radius parameter obtained using this prescription is 1.65 fm, which
is within 10$\%$ of the radius parameter we have used to reproduce the
experimental data. However, it may be mentioned here that the present
simplified approach may not lead to realistic values of $r_\circ$ for
larger values of compound nucleus spin, which may limit the use
of this simple approach for heavier compound systems at higher spins. 
In conclusion, the present work demonstrates that
a modified radius parameter (r$_{\circ}$) alone  may be sufficient to 
reproduce the deformations for lighter systems
(A $\sim$ 30).  However, it will be interesting to explore the validity of this
simple approach in explaining the light charged particle data for  
heavier systems at higher excitations.

\acknowledgements

The  authors  thank  the Pelletron operating staff for smooth running of
the machine and Mr. D. C. Ephraim of T.I.F.R., for making the targets.

\nopagebreak

\end{document}